\documentclass[12pt]{article}

\usepackage{amsmath} 
\usepackage{amssymb}
\usepackage{hhline}
\usepackage[scale={.75,.75}]{geometry}
\usepackage{url}
\usepackage{lscape}
\usepackage{caption}
\usepackage[numbers,sort&compress]{natbib}
\usepackage{epsfig}
\usepackage{multirow} 
\usepackage{color}
\usepackage{verbatim}
\usepackage{nicefrac}
\usepackage{upgreek}
\usepackage{bbm}
\usepackage{setspace}
\usepackage{pstricks}
\usepackage{psfrag}
\usepackage{color}

\usepackage{hyperref}
\usepackage{upgreek}
\usepackage{bbm}
\usepackage{graphicx}
\usepackage{psfrag}
\usepackage{verbatim}

\newcommand{\slashed}[1]{\displaystyle{\not}{#1}}

\begin{document}


\title{\vspace{-2.5cm} 
\begin{flushright}
\vspace{-0.4cm}
{\scriptsize \tt TUM-HEP-939/14}  
\end{flushright}
{\bf Probing leptogenesis with GeV-scale sterile neutrinos at LHCb and BELLE II
}
}
\author{Laurent Canetti, Marco Drewes$^1$, Bj\"orn Garbrecht$^1$\\ 
\scriptsize{$^1$Physik Department T70, Technische Universit\"at M\"unchen, James Franck Stra\ss e 1, D-85748 Garching, Germany}
\vspace{-0.7cm}
}
\date{}

\maketitle

\begin{abstract}
  \noindent
  We show that existing laboratory experiments have the potential to unveil the origin of matter by probing leptogenesis in the type-I seesaw model with three right-handed neutrinos and Majorana masses in the GeV range.
The baryon asymmetry is generated by CP-violating flavour oscillations during the production of the right-handed neutrinos. 
In contrast to the case with only two right-handed neutrinos, no degeneracy in the Majorana masses is required.
The right-handed neutrinos can be found in meson decays at BELLE II and LHCb.
\end{abstract}


\section{Introduction}

All matter particles in the Standard Model 
(SM) except neutrinos have been observed with both, left-handed (LH) and right-handed (RH) chirality.
If RH neutrinos exist, they can explain several phenomena which cannot be understood in the framework of the SM, for a review see e.g. Refs.~\cite{Abazajian:2012ys,Drewes:2013gca}.
In particular, they give neutrinos masses via the seesaw mechanism \cite{Minkowski:1977sc,GellMann:seesaw,Mohapatra:1979ia,Yanagida:1980xy} and can at the same time generate the baryon asymmetry of the universe (BAU) via leptogenesis \cite{Fukugita:1986hr}.
Leptogenesis with $n=2$ RH neutrinos has been studied in detail, and it was found that the BAU can only be explained if their masses are either very heavy \cite{Davidson:2002qv,Antusch:2009gn,Racker:2012vw} or degenerate \footnote{The $\nu$MSM-scenario  proposed in \cite{Asaka:2005an,Asaka:2005pn} involves three sterile flavours, but one of them is required to be a Dark Matter candidate and couples to weakly to contribute to the seesaw and leptogenesis, see \cite{Boyarsky:2009ix,Canetti:2012kh} for details.}, see e.g. \cite{Pilaftsis:1997jf,Pilaftsis:2003gt} and \cite{Asaka:2005pn,Canetti:2010aw,Asaka:2011wq,Canetti:2012kh,Canetti:2012vf,Shuve:2014zua,Garbrecht:2014bfa}.
In the former case the new particles are too heavy to be seen in any experiment. In the latter case it was found that their interaction strengths, characterized by mixing angles with ordinary neutrinos, 
are generally too feeble to give measurable branching ratios in existing experiments \cite{Kersten:2007vk,Ibarra:2011xn,Canetti:2012vf,Canetti:2012kh}. They could be found in dedicated future experiments \cite{Bonivento:2013jag} using present day technology \cite{Canetti:2012vf,Canetti:2012kh}.
In this work we show that both shortcomings, the ``tuned'' mass degeneracy and suppressed production rates at colliders, are specific to the scenario with $n=2$ and can be avoided if three or more RH neutrinos participate in leptogenesis even if there is no other physics beyond the SM.

The present article is organized as follows.
In Sec.~\ref{secmodel} we recapitulate the type-I seesaw model and introduce our notation. 
In Sec.~\ref{sec:leptogenesis} we briefly summarize the mechanism of leptogenesis via neutrino oscillations.
In Sec.~\ref{parameter:scan} we identify a range of sterile neutrino masses and mixings that can explain the observed BAU.
We discuss the perspectives to explore this region experimentally in Sec.~\ref{exp:persp}.
In Sec.~\ref{TuningBlaBlaBla} we comment on the dependence of the baryogenesis region on the choice of model parameters.
We draw conclusions in section \ref{sec:conclusions}.

\section{The model}\label{secmodel}
We consider the most general renormalizes Lagrangian in Minkowski space that only contains SM fields and RH neutrinos $\nu_R$, 
\begin{eqnarray}
	\label{L}
	\mathcal{L} &=&\mathcal{L}_{SM}+ 
	i \overline{\nu_{R}}\slashed{\partial}\nu_{R}-
	\overline{l_{L}}F\nu_{R}\tilde{\Phi} -
	\tilde{\Phi}^{\dagger}\overline{\nu_{R}}F^{\dagger}l_L \nonumber\\ 
	&&-{\rm \frac{1}{2}}(\overline{\nu_R^c}M_{M}\nu_{R} 
	+\overline{\nu_{R}}M_{M}^{\dagger}\nu^c_{R}). 
	\end{eqnarray}
Here flavour and isospin indices are suppressed.
$\mathcal{L}_{SM}$ is the SM Lagrangian, 
$l_{L}=(\nu_{L},e_{L})^{T}$ are the left handed SM lepton doublets, $\Phi$ is the Higgs doublet and 
$\tilde{\Phi}=\epsilon\Phi^*$, where $\epsilon$ is the antisymmetric $SU(2)$ tensor. 
$F$ is a matrix of Yukawa couplings and $M_{M}$ a Majorana mass term for $\nu_{R}$ with $\nu_R^c=C\overline{\nu_R}^T$. 
The charge conjugation matrix is $C=i\gamma_2\gamma_0$.
For $n$ flavours of $\nu_R$, the eigenvalues of $M_M$ introduce $n$ new mass scales in nature. 
In analogy with the LH sector we consider the case of $n=3$ flavours of RH neutrinos. 
This is the minimal number required to generate three non-zero light neutrino masses. 
We work in a flavour basis where $M_M={\rm diag}(M_1,M_2,M_3)$.
For $M_I\gg 1$  eV one observes two distinct sets of mass eigenstates, which we represent by flavour vectors of Majorana spinors $\upnu$ and $N$.
The elements $\upnu_i$ of the flavour vector 
\begin{equation}
\upnu=V_\nu^{\dagger}\nu_L-U_\nu^{\dagger}\theta\nu_{R}^c +{\rm c.c.}
\end{equation}
are mainly superpositions of the ``active'' SU(2) doublet states $\nu_L$ and have light masses $\sim m_\nu=-\theta M_M \theta^T\ll M_M$. The elements $N_I$ of 
\begin{equation}
N=V_N^\dagger\nu_R+\Theta^{T}\nu_{L}^{c} +{\rm c.c.}
\end{equation}
are mainly superpositions of the ``sterile'' singlet states $\nu_R$ and have masses of the order of $M_I$. 
Here c.c. stands for the $c$-conjugation defined above, $\Theta\ll \mathbbm{1}$ is the mixing matrix between active and sterile neutrinos and $\theta\equiv\Theta U_N^T$. $V_\nu$ is the usual neutrino mixing matrix and $U_\nu$ its unitary part, $V_N$ and $U_N$ are their equivalents in the sterile sector.
To be precise: $V_\nu\equiv (\mathbbm{1}-\frac{1}{2}\theta\theta^{\dagger})U_\nu$
with  $\theta\equiv m_D M_M^{-1}$,  $m_D\equiv Fv$ and the temperature dependent Higgs field expectation value $v$ ($v=174$ GeV at temperature $T=0$).
The unitary matrices $U_\nu$ and $U_N$ diagonalize the mass matrices 
\begin{eqnarray}
m_\nu\simeq-\theta M_M \theta^T = -\frac{1}{v^2}F M_M^{-1}F^T \ {\rm and} \label{activemass}\\
M_N=M_M + \frac{1}{2}\big(\theta^{\dagger} \theta M_M + M_M^T \theta^T \theta^{*}\big), 
\end{eqnarray}
respectively.
Experimentally the magnitude of the $M_I$ is almost unconstrained, as neutrino oscillation experiments at energies $E\ll M_I$ only involve the light states $\nu_L$ and probe the eigenvalues of the combination $m_\nu m_\nu^\dagger$, 
which can be matched for any choice of $M_I$ between sub-eV values \cite{deGouvea:2005er} and the scale of grand unification if the $F_{\alpha I}$ are chosen appropriately.
The phenomenological implications for different choices of these parameters can be extremely different, see e.g. Refs.~\cite{Drewes:2013gca,Abazajian:2012ys} for a summary.
Confirmation of the model (\ref{L}) requires the masses of the new states $N_I$ to be within reach of experiments.

For $n=3$ the Lagrangian (\ref{L}) contains $18$ new physical parameters. 
We use the Casas-Ibarra parametrization  \cite{Casas:2001sr} for the Yukawa matrices
\begin{equation}
F=\frac{1}{v}U_\nu\sqrt{m_\nu^{\rm diag}}\mathcal{R}\sqrt{M_M}.
\end{equation}
Here $m_\nu^{\rm diag}=U_\nu^\dagger m_\nu U_\nu^*={\rm diag}(m_1,m_2,m_3)$ and $\mathcal{R}$ is a matrix with $\mathcal{R}^T\mathcal{R}=\mathbbm{1}$ that can be parametrized by complex mixing angles $\omega_{ij}$ as 
\begin{equation}
\mathcal{R}=\mathcal{R}^{(23)}\mathcal{R}^{(13)}\mathcal{R}^{(12)}\,.
\end{equation}
with the non-zero elements
\begin{eqnarray}
\mathcal{R}^{(ij)}_{ii}&=&\mathcal{R}^{(ij)}_{jj}=\cos\omega_{ij} ,\nonumber\\
\mathcal{R}^{(ij)}_{ij}=\sin\omega_{ij} \ &,& \
\mathcal{R}^{(ij)}_{ji}=-\sin\omega_{ij} \ , \ 
\mathcal{R}^{(ij)}_{kk}\underset{k\not=i,j}{=}1.\nonumber
\end{eqnarray}
In the flavour basis where the charged lepton Yukawa couplings are diagonal the matrix $U_\nu$ can be parameterized as
\begin{equation}
U_\nu=V^{(23)}U_{\delta}V^{(13)}U_{-\delta}V^{(12)}{\rm diag}(e^{i\alpha_1 /2},e^{i\alpha_2 /2},1) \nonumber
\end{equation}
with $U_{\pm\delta}={\rm diag}(e^{\mp i\delta/2},1,e^{\pm i\delta/2})$ and 
the non-zero entries of the matrices $V$ are given by
\begin{eqnarray}
V^{(ij)}_{ii}&=&V^{(ij)}_{jj}=\cos\uptheta_{ij} , \nonumber\\
V^{(ij)}_{ij}=\sin\uptheta_{ij} \ &,& \
V^{(ij)}_{ji}=-\sin\uptheta_{ij} \ , \ 
V^{(ij)}_ {kk}\underset{k\not=i,j}{=}1 \nonumber
\end{eqnarray}
where 
$\uptheta_{ij}$ are the mixing angles
amongst the active leptons, and $\alpha_1$, $\alpha_2$ and $\delta$ are the $CP$-violating phases.
This allows to directly encode all constrains from neutrino oscillation experiments in $U_\nu$ and $m_\nu^{\rm diag}$.
For $n=3$ sterile flavours there are three complex ``Euler angles'' $\omega_{ij}$ in $\mathcal{R}$, while for $n=2$ there would be only one such angle $\omega$.
We study the perspectives to find the new states $N_I$ in laboratory experiments, in particular LHCb and BELLE II, and focus on the mass range $M_I<5$ GeV.

\section{Leptogenesis}\label{sec:leptogenesis}  
Leptogenesis can explain the matter-antimatter asymmetry in the universe by CP-violating interactions of the $N_I$ in the primordial plasma. 
This requires a deviation from thermal equilibrium \cite{Sakharov:1967dj}, hence it can occur either during the production \cite{Akhmedov:1998qx,Asaka:2005pn} or the freezeout and decay \cite{Fukugita:1986hr} of $N_I$ in the early universe. 
We focus on the scenario where the BAU is generated during $N_I$ production, which is often referred to as \emph{baryogenesis from neutrino oscillations}.  
Due to their feeble interactions the $N_I$ are not produced in significant amounts during cosmic inflation and reheating \cite{Bezrukov:2008ut} and have in good approximation zero abundance at the onset of the radiation dominated era.
They are then produced thermally from the primordial plasma.
The combination of oscillations and CP-violating scatterings during this nonequilibrium process can generate a matter-antimatter asymmetry.
For the parameters we consider the CP-violation contained in $F$ typically acts most efficiently when the primordial plasma has a temperature $T\sim10^5$ GeV, where sphaleron processes rapidly violate baryon number $B$ \cite{Kuzmin:1985mm}, 
but conserve $B-L$.
The violation of total lepton $L$ number is suppressed by $M_I/T\ll 1$. However, there can be significant asymmetries $L_\alpha$ in the individual flavours, with $L=\sum_\alpha L_\alpha$. 
For $M_I/T \ll 1$ the helicity states of the Majorana fields $N_I$ effectively act as "particles" and "antiparticles", and one can assign approximately conserved lepton numbers $L_I$ to the sterile flavours.
The sum $L+\sum_I L_I$ is conserved in very good approximations at the temperatures $T\gg M_I$ we consider.
The violation of the individual $L_\alpha$ occurs at order $\mathcal{O}[F^4]$, while $L$-violation occurs at order $\mathcal{O}[F^6]$ via flavour dependent scatterings.
These convert a part $\delta L$ of the lepton asymmetry 
into the $L_I$, where it is hidden from the sphaleron processes that 
partly transfer the remaining net asymmetry $-\delta L$ into $B$. 
The $L_\alpha$ and $B$ get washed out once the $N_I$ come into equilibrium. If this process is incomplete at the time of sphaleron freezeout at $T=T_{\rm sph}\sim 130-140$ GeV \cite{Burnier:2005hp,D'Onofrio:2012jk,D'Onofrio:2014kta}, after which $B$ is conserved, then a net $B\neq 0$ remains protected from further washout at lower temperatures. This mechanism is explained in more detail in Refs.~\cite{Boyarsky:2009ix,Canetti:2012kh,Drewes:2012ma,Drewes:2013gca,Khoze:2013oga,Shuve:2014zua}.

The rate of thermal $N_I$ production is given by $\Gamma_I=(F^\dagger F)_{II}\gamma_{av}T$. The quantity $\gamma_{av}$ is a numerical coefficient that depends on $M_I/T$ and has to be calculated in thermal field theory. 
The asymmetry is bigger if generated at earlier times, hence larger couplings $F$ give a larger BAU. 
On the other hand, larger  $F$ also imply larger washout rates $\Gamma_\alpha=(F F^\dagger)_{\alpha\alpha}\tilde{\gamma}_{av}T$.
With $\tilde{\gamma}_{av}\simeq \gamma_{av}$ it is crucial that the Yukawa interactions $F$ are large enough to generate significant lepton asymmetries $L_\alpha$ at $T\gg T_{\rm sph}$,  
but small enough to prevent a complete washout of all $L_\alpha$ before $T=T_{\rm sph}$.
This is most easily achieved if individual elements $F_{\alpha I}$ are sufficiently different in size  that one active flavour $\alpha$ couples much more weakly to the $N_I$ than the other two, leading to a flavour asymmetric washout that allows the asymmetry in that flavour to survive until $T=T_{\rm sph}$.  
For the sake of definiteness we assume in the following that this is the electron flavour, i.e.
\begin{equation}\label{electronsmall}
\Gamma_e\ll \Gamma_{\mu,\tau}.
\end{equation}
In the minimal scenario with $n=2$ this is difficult to achieve because there is only one complex angle $\omega$ in $\mathcal{R}$. The strengths of the active-sterile couplings in all flavours are tied together, as they are essentially governed by just one parameter ${\rm Im}\omega$ \cite{Gorbunov:2007ak,Shaposhnikov:2008pf,Asaka:2011pb,Shuve:2014zua,Gorbunov:2013dta}. 
This generally leads to very small baryon asymmetries because a large asymmetry generation at $T\gg T_{\rm sph}$ necessarily implies a large washout for all flavours at $T\gtrsim T_{\rm sph}$, and the observed BAU can only be explained if it is resonantly enhanced by a degeneracy in the masses at the level $<10^{-3}$ \cite{Canetti:2012kh,Shuve:2014zua}. 
At the same time the non-observation of $\mu\rightarrow e \gamma$ implies strong upper bounds on the interactions of all flavours, which makes a detection at colliders difficult \cite{Ibarra:2011xn}.
The situation changes drastically in the $n=3$ scenario.
The reason is that in this case there are three complex angles $\omega_{ij}$ in $\mathcal{R}$. This enlarged parameter space contains considerable regions in which $\Gamma_e\ll \Gamma_{\mu,\tau}$.

In the following we exclude the mass degenerate case $|M_I-M_J|\lesssim \Gamma_I$ from our analysis, which requires a more sophisticated treatment of flavour oscillations \cite{Garny:2011hg,Garbrecht:2011aw,Iso:2013lba,Dev:2014laa,Iso:2014afa,Hohenegger:2014cpa} and makes up only a small fraction of the parameter space. 
This allows to approximate $V_\nu=U_\nu$ and $U_N=\mathbbm{1}$.
The lepton charge $q_\alpha$ in the flavour $\alpha$ at $T\gg T_{\rm sph}$ can be estimated as
\begin{eqnarray}
\label{flavoured:asymmetries}
\frac{q_{\alpha}}{s}
&\approx&
-\sum\limits_{\overset{\beta}{J\not=I}}
i
\frac
{
F_{\alpha I}F^\dagger_{I\beta}F_{\beta J} F^\dagger_{J \alpha}
-F^*_{\alpha I}F^T_{I \beta}F^*_{\beta J}F^T_{J \alpha}
}
{
{\rm sign}(M_I^2-M_J^2)
}\nonumber\\
&&\times
\left(\frac{m_{\rm Pl}^2}{|M_I^2-M_J^2|}\right)^{\frac 23} 1.2\times 10^{-4}\gamma_{\rm av}^2\
.\end{eqnarray}
Here $s$ is the entropy density and $m_{\rm Pl}$ the Planck mass. 
We define charges $q_I$ in the $N_I$ via their helicity states.
The expression (\ref{flavoured:asymmetries}) has been derived in \cite{Drewes:2012ma} within the framework of the nonequilibrium quantum field theory approach to leptogenesis \cite{Buchmuller:2000nd,DeSimone:2007rw,Garny:2009rv,Garny:2009qn,Anisimov:2010aq,Garny:2010nj,Beneke:2010wd,Garny:2010nz,Garbrecht:2010sz,Beneke:2010dz,Anisimov:2010dk,Fidler:2011yq,Garbrecht:2012qv,Garbrecht:2011aw,Garny:2011hg,Drewes:2012ma,Frossard:2012pc} and allows to systematically include the different quantum and thermodynamical effects that affect the asymmetry generation in a controlled approximation scheme.
Up to numerical coefficients it agrees with an expression found in \cite{Asaka:2005pn} in the framework of \emph{density matrix equations} \cite{Sigl:1992fn}.
In this work we want to show, as a proof of principle, that leptogenesis is possible with experimentally accessible $N_I$.
For this purpose we restrict ourselves to the case \begin{equation}\label{assumption}
\Gamma_\mu/H, \Gamma_\tau/H > 1  \  {\rm at} \ T=T_{\rm sph}
.\end{equation}
Here $H=\sqrt{8\pi^3g_*/90}T^2/m_{\rm Pl}$ is the Hubble rate and $g_*$ the number of relativistic degrees of freedom in the primordial plasma.
We denote charge densities in the $\mu$-leptons, $\tau$-leptons and the $N_I$ after their equilibration by $q_{\mu,\tau}^{\rm W}$ and $q_I^W$.
Equilibration of the chemical potentials gives the relation 
$2 q^{\rm W}_{\mu}=2 q^{\rm W}_{\tau}=q^{\rm W}_{I}$, where the factor $2$ counts the SU(2) doublet components.
 Lepton number conservation implies
$q^{\rm W}_{\mu}+q^{\rm W}_{\tau}+q^{\rm W}_{1}+q^{\rm W}_{2}+q^{\rm W}_{3}=q_\mu+q_\tau=-q_e$, hence $q^{\rm W}_{\mu}+q^{\rm W}_{\tau}=-(4/7)q_e$ is the part of the asymmetry that is canceled before sphaleron freezeout.
Taking account of the $q_e$ washout and the sphaleron conversion factor \cite{Khlebnikov:1988sr,Laine:1999wv}, the BAU can be obtained as 
\begin{equation}\label{BAU}
\frac{q_B}{s}\simeq -\frac{28}{79}\frac{q_e}{s}\frac{3}{7}e^{-\Gamma_e/H}
.\end{equation}

\section{Parameter scan}\label{parameter:scan}
We aim to identify the overlap between the parameter region where baryogenesis is possible and the experimentally accessible range of masses and mixings. 
We expect that the baryogenesis region for given $M_I$ extends between a lower and upper bound on the angles $|\Theta_{\alpha I}|$.
The lower bound comes from the requirement to produce enough asymmetry (\ref{flavoured:asymmetries}), the upper bound from the requirement to keep at least one of the $\Gamma_\alpha$ small enough to prevent a complete washout of the flavored asymmetries before sphaleron freezeout.
To determine the overlap with the experimentally accessible region, we perform a parameter scan to identify the largest mixing $|\Theta_{\alpha I}|$ for which the observed BAU can be explained.
We compare it to current experimental upper bounds and to the expected future sensitivity. 
For the sake of definiteness we study the the mixing between $N_2$ and $\nu_\mu$, which has e.g. been probed at LHCb. 
We have checked that the perspectives for $N_1$ and $N_3$ are similar.
We restrict ourselves to the fraction of the parameter space where
(\ref{assumption}) is fulfilled and (\ref{BAU}) can be used to compute the BAU.
This is precisely the region where one can expect large $|\Theta_{\mu I}|$.

We choose the Casas-Ibarra parametrization \cite{Casas:2001sr} 
defined above and fix $M_1=1$ GeV, $M_3=3$ GeV, $m_1 = 2.5\times 10^{-3}$ eV, 
$m_2 = 9.05 \times 10^{-3}$ eV and
$m_3 = 5 \times 10^{-2}$ eV.
We fix all other known neutrino parameters according to the global fits given in \cite{Capozzi:2013csa} and vary $M_2$.
This approach is valid as long as radiative corrections $\delta m_\nu$ to $m_\nu$ are small \cite{AristizabalSierra:2011mn}. We restrict ourselves to the region where the effect of the one-loop correction \cite{Pilaftsis:1991ug}
\begin{eqnarray}
\lefteqn{\delta (m_\nu)_{\alpha \beta} = \sum_I\frac{1}{(4\pi)^2}F_{\alpha I}M_IF^T_{I \beta}}\nonumber \\
&&\times\left(
\frac{3\ln[(M_I/m_Z)^2]}{(M_I/m_Z)^2-1}
+\frac{\ln[(M_I/m_H)^2]}{(M_I/m_H)^2-1}
\right)\label{RadiativeCorrections}
\end{eqnarray}
on the observed $\delta m^2_{\rm sol}$ and $\delta m^2_{\rm atm}$ is smaller than the $2 \sigma$ uncertainty quoted in \cite{Capozzi:2013csa}.
For each choice of $M_2$ we perform a Monte Carlo scan of $5\times10^8$ points over the Majorana phases $\alpha_1$ and $\alpha_2$, the Dirac phase $\delta$ and all three complex mixing angles $\omega_{ij}$ to identify the parameter region where the observed BAU is explained. For all ${\rm Im}\omega_{ij}$ we scan the interval $[-5,5]$, the dependence on all other parameters is periodic.
The parameter $\gamma_{av}$ can be determined from the results for $\Gamma_I$ found in Refs.~\cite{Besak:2012qm,Besak:2012qm,Garbrecht:2013bia,Bodeker:2014hqa,Laine:2013lka}, which slightly differ from each other. 
We use the results from Ref.~\cite{Garbrecht:2013bia}, which imply $\gamma_{av}=0.015$ at $T=T_{\rm sph}$ and $\gamma_{av}=0.013$ at $T=10^5$ GeV for $M_I\ll T_{\rm sph}$.

\begin{figure}[h!]
\centering
\includegraphics[width=12cm]{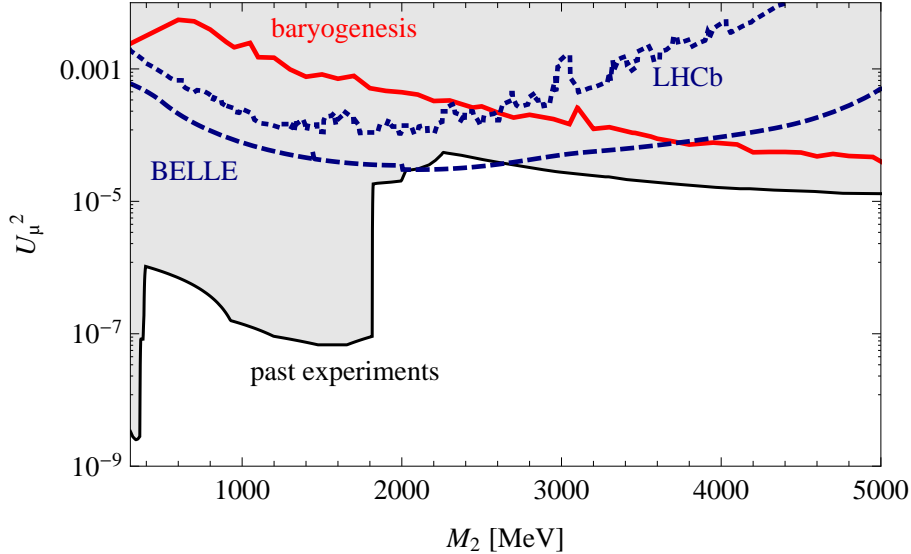}
\caption{The red line shows the maximal mixing $|\Theta_{\mu 2}|^2$ we found consistent with baryogenesis, i.e. below the line there exist parameter choices for which the observed BAU can be generated.
The scatter is a result of the Monte Carlo method and not physical. It indicates that we have not found the global maxima, but the density of valid points decreases rapidly for larger $|\Theta_{\mu 2}|^2$.
The gray area represents bounds from the past experiments PS191 \cite{Bernardi:1987ek}, NuTeV \cite{Vaitaitis:1999wq} (both re-analyzed in \cite{Ruchayskiy:2011aa}), NA3 \cite{Badier:1985wg}, CHARMII \cite{Vilain:1994vg} and DELPHI \cite{Abreu:1996pa} (as given in \cite{Atre:2009rg}). They are stronger than those from violation of lepton universality \cite{Abada:2012mc,Abada:2013aba,Basso:2013jka,Endo:2014hza}, see \cite{Atre:2009rg,Drewes:2013gca} for a discussion of other experimental constraints. The blue lines indicate the current bounds from LHCb \cite{Aaij:2014aba} (dotted) and BELLE \cite{Liventsev:2013zz} (dashed), which will improve in the future. 
\label{bounds}
}
\end{figure}

The BAU can be measured in different ways, see e.g. Ref.~\cite{Canetti:2012zc} for a review. 
Here we use the value $q_B/s=8.58\times10^{-11}$ inferred from the Planck data \cite{Ade:2013zuv}.
We accept a point if (\ref{BAU}) gives at least this value, as the asymmetry can always be reduced by varying the CP-violating phases.
We also require each point to be consistent with bounds on lepton flavour violation on the low-scale seesaw \cite{Ibarra:2011xn,Canetti:2013qna} from the rare decay $\mu\rightarrow e\gamma$ in the MEG experiment \cite{Adam:2013mnn} and the limits on neutrinoless double $\beta$-decay 
\cite{Bezrukov:2005mx,LopezPavon:2012zg,Asaka:2013jfa,Merle:2013ibc,Girardi:2013zra,Ibarra:2011xn,LopezPavon:2012zg} obtained by the GERDA experiment \cite{Agostini:2013mzu}.
These are currently the strongest bounds on lepton flavour violation and total lepton number violation, respectively \cite{Ibarra:2011xn,Dinh:2012bp,Lello:2012in}.
For the branching ratio of the decay $\mu\rightarrow e\gamma$ we use the expression \cite{Cheng:1980tp,Bilenky:1977du}
\begin{eqnarray}\label{BranchingRatio}
B(\mu\rightarrow e \gamma) = \frac{\Gamma(\mu\rightarrow e \gamma)}{\Gamma(\mu\rightarrow e \nu_\mu \bar{\nu}_e)} = \frac{3\alpha_{\rm em}}{32\pi}|R|^2
\end{eqnarray}
with
\begin{eqnarray}
R&=&\sum_{i} (V_\nu^*)_{\mu i} (V_\nu)_{e i}
G\left(\frac{m_i^2}{M_W^2}\right)
+\sum_I\Theta^*_{\mu I} \Theta_{e I}G\left(\frac{M_I^2}{M_W^2}\right)\nonumber
\end{eqnarray}
and the loop function
\begin{equation}
G(x)=\frac{10-43x+78 x^2-49 x^3 + 4 x^4 + 18 x^3 \log(x)}{3 (x - 1)^4}.\nonumber
\end{equation}
and demand $B(\mu\rightarrow e \gamma)<5.7\times10^{-13}$ \cite{Adam:2013mnn}.
To be consistent with the non-observation of neutrinoless double $\beta$-decay 
we impose the bound $m_{ee}<0.2$ eV \cite{Agostini:2013mzu} on the \emph{effective Majorana mass}  
\begin{equation}\label{mee}
m_{ee}=\left|
\sum_i (U_\nu)_{ei}^2m_i + \sum_I \Theta_{eI}^2M_I f_A(M_I)
\right|.
\end{equation}
Here $f_A=(M_A/M_I)^2f_A$, $M_A\simeq 0.9$ GeV and $f_A=0.079$ for $^{76}\rm{Ge}$.
Finally, the lifetime of the RH neutrinos is constrained by the requirement that they decay faster than about $0.1$s. Otherwise they would affect the formation of light elements in big bang nucleosynthesis (BBN) in the early universe.
To evaluate this criterion, we used the decay rates given in \cite{Gorbunov:2007ak} and \cite{Canetti:2012kh}.
The results of the scan are displayed in FIG.~\ref{bounds}.  

\section{Experimental perspectives}\label{exp:persp}
From an experimental viewpoint there are three qualitatively different mass regions. In all regions $N_2$ can be produced in real $W$ and $Z$-boson decays.
In region $i)$ with $M_2< 2$ GeV $N_I$ can also be produced in the decays $D$-mesons and $B$-mesons, in region $ii)$ with $2 {\rm GeV} < M_2 < 5$ GeV $N_2$ can be produced in the decay of $B$-mesons. 
In region $iii)$ with $M_2>5$ GeV $N_2$ is too heavy to be produced in meson decays.
In FIG.~\ref{bounds} we show the largest $|\Theta_{\mu 2}|^2$ we found that can lead to baryogenesis as a function of $M_2$ in regions $i)$ and $ii)$. 
This need not be an absolute upper bound, as there may be parameter choices outside the region we scanned that yield an even larger mixing. 
However, the precise value of the largest mixing that would be consistent with baryogenesis is practically not relevant in the mass range we consider because the values we found lie considerably above the experimental limits.
This is also the reason why the uncertainties in our calculation in the coefficient $\gamma_{av}$, the temperature dependence of the sphaleron rate near $T_{\rm sph}$ \cite{D'Onofrio:2014kta} and from the neglected momentum dependence \cite{Asaka:2011wq} in (\ref{BAU}) do not affect our conclusion.

When comparing our bounds to limits inferred from experimental searches, it should be kept in mind that experiments usually quote limits that are obtained under the assumption that there is only $n=1$ RH neutrino.
For $n=2$, leptogenesis with $M_I$ in the GeV range requires that both masses are degenerate with a common mass $M=(M_1+M_2)/2$ and a splitting $|M_1-M_2|/M<10^{-3}$ that is too small to be resolved experimentally \cite{Canetti:2012kh}. 
Hence, one can convert the bounds quoted by experiments into constraints $U_\mu^2=\sum_I |\Theta_{\mu I}|^2$, as both $N_I$ lead to experimental signatures at the same mass $M$ \cite{Ruchayskiy:2011aa}.
For $n=3$ the relation between the measured branching ratios and $|\Theta_{\alpha I}|^2$ is more complicated; it depends on several parameters
and cannot be displayed easily. In particular, there is no need for a mass degeneracy, so the individual $|\Theta_{\mu I}|$ will lead to signals at different $M_I$. Since $U_\mu^2>|\Theta_{\mu 2}|^2$, comparing the theoretical $|\Theta_{\mu 2}|^2$ to the bound on $U_\mu^2$ should therefore be regarded as a conservative estimate of the experimental perspectives.

FIG.~\ref{bounds} shows that any experiment which improves the known bounds  has the potential to discover the $N_I$ responsible for baryogenesis.    
The most stringent bounds from past experiments have been summarized in \cite{Atre:2009rg,Ruchayskiy:2011aa,Canetti:2012kh,Asaka:2012bb}. 
In region $i)$ these have already deeply entered into the cosmologically relevant parameter space, they impose $U_\mu^2< 10^{-9}-10^{-4.9}$ for $M_I$ below the Kaon mass and $U_\mu^2<10^{-7.5}-10^{-6.3}$ below the masses of $D$-mesons. No existing experiment has the potential to improve these bounds. 
However, the SHiP experiment proposed in \cite{Bonivento:2013jag,Alekhin:2015oba,Anelli:2015pba} would be able to probe a significantly larger fraction of the cosmologically interesting parameter space down to $U_\mu^2\sim 10^{-9}$ up to the D-meson mass. 
In region $ii)$ the strongest bounds come from the search for heavy neutral leptons in $Z$-boson decays in the DELPHI experiment \cite{Abreu:1996pa}, from LHCb \cite{Aaij:2014aba} and from the BELLE experiment \cite{Liventsev:2013zz}. 
All of them have already entered the parameter region where baryogenesis is possible.
The published LHCb bounds are below those from DELPHI, but LHCb will continue to take data after the upgrade to 14 TeV. 
The displayed bounds were obtained with an integrated luminosity $L_{\rm int}=3 {\rm fb}^{-1}$ from the process $B^-\rightarrow \pi^+\mu^-\mu^-$. 
The branching ratio for this decay is $\propto |\Theta_{\mu 2}|^4$, hence the bound on $|\Theta_{\mu 2}|$ under ideal conditions improves as $\propto\sqrt{L_{\rm int}}$.
With the anticipated  $L_{\rm int}=50 {\rm fb}^{-1}$ the bound will improve by a factor $\sim(2\times 3/50)^{1/2}\sim 6$ in the region where the background is low, just enough to outperform DELPHI and current BELLE bounds.
The factor $2$ comes from the enhanced production of $B$-mesons at the $14$ TeV LHC. Further improvement can be achieved by analysing additional processes, e.g. the leptonic and semi-leptonic decays listed in \cite{Gorbunov:2007ak,Atre:2009rg,Canetti:2012kh}. These include decays with other mesons or electrons in the final state. 
Though heavier mesons in the final state kinematically restrict the range of accessible $M_I$ and the reconstruction efficiency for electrons is lower, in combination these could significantly improve the bounds at least in some part of the mass region below $5$ GeV.
The displayed bounds from BELLE are stronger because they include a number of different processes. 
The increase in $L_{\rm int}$ alone will make BELLE II an order of magnitude more sensitive to $U_\mu^2$ than BELLE. Further improvement is possible if data from more channels is used in the analysis.
This includes ``peak searches'' for missing four-momentum, which can be competitive in spite of a $\sim 10^3$ times worse reconstruction efficiency because the branching ratio scales $\propto U_\alpha^2$, and the sensitivity improves linearly with $L_{\rm int}$.
Though not shown in FIG.~\ref{bounds}, also SHiP could probe region $ii)$. The expected sensitivity has not been calculated yet.
Region $iii)$ could be probed in gauge boson decays, but no dedicated experiment of this kind is currently planned.

To see if any realistic experiment could completely rule out this baryogenesis scenario one has to determine the lower bound on the mixing. 
Our formula (\ref{BAU}) to calculate the BAU can only be applied in the parameter region where (\ref{assumption}) is fulfilled, which is also the experimentally most easily accessible region due to the large mixing with $\nu_\mu$. 
This strategy does not allow to find the strict lower bound on $|\Theta_{\mu 2}|^2$ for which the BAU can be generated.
A complete analysis requires the numerical solution of quantum kinetic equations with coefficients that can be derived systematically in thermal field theory, which we postpone  to future work. 
The smallest mixings we find for $0.5 {\rm GeV} < M_2 < 5$ GeV 
within the range of validity of (\ref{assumption}) and in absence of mass degeneracies are $U_\mu^2\lesssim 10^{-10}$, which is an order of magnitude below the anticipated SHiP sensitivity.

A strict lower bound on the mixing can in principle be derived using data from neutrino oscillation experiments independently of leptogenesis. 
For $n=2$ degenerate sterile neutrinos a minimal active-sterile mixing $U_\mu^2$ can be derived from that \cite{Asaka:2011pb}.
For the scenario considered here with $n=3$, the situation is more complicated. 
In general, there is no lower bound on any individual $|\Theta_{\alpha I}|^2$ \cite{Gorbunov:2013dta}. For instance, $\Theta_{\mu 2}$ can be set to zero by some choices of $\omega_{23}$.\footnote{The required value for $\omega_{23}$ can be found analytically, but the expression is lengthy and by itself not illuminating.}
It is not possible to suppress the mixing of all $N_I$ if one wants to explain the observed neutrino oscillations, i.e. $U_\mu^2$ must be non-zero and at least in principle has a lower bound. 
If we, for instance, require all $M_I$ to lie in region $i)$, then we numerically find that at least one $|\Theta_{\mu I}|^2$ must be larger than $\sim10^{-13}$ if the sum of neutrino masses is near the cosmological limit of $0.23$ eV \cite{Ade:2013zuv} or $\sim 10^{-12}$ if the lightest neutrino is massless (both for normal hierarchy).
However, if only one of the $N_I$ has a mass in region $i)$ and the other two are heavier, then the mixing of the light one can be arbitrarily small, which seems discouraging for experimental searches. Note, however, that the considerations in this paragraph are independent of leptogenesis. It could be that the requirement to explain the BAU with a non-degenerate mass spectrum imposes a stronger lower bound on the active sterile mixing. This seems likely because if one of the $N_I$ has an extremely suppressed mixing with active flavours, then it essentially decouples and one effectively recovers a model with $n=2$ as far as baryogenesis and neutrino masses are concerned, in which case such constraint exists \cite{Canetti:2012kh}.

To confirm that the RH neutrinos are indeed responsible for baryogenesis in the early universe, one would also have to measure the CP-violation in their interactions. Three of the six CP-violating parameters in the model can be identified with the Dirac and Majorana phases $\delta$ and $\alpha_1$, $\alpha_2$ in the neutrino mixing matrix $V_\nu$, which in principle experimentally accessible. 
With present day technology the remaining CP-violation in the sterile sector could only be probed in the parameter region where the mass spectrum is degenerate \cite{Cvetic:2014nla}.

\section{Parameter space measures}\label{TuningBlaBlaBla}
In the context of model building it is common to pose the questions how "natural" or "generic" the parameter choice that leads to interesting phenomena is.
While such considerations provide useful guidelines for theorists in face of the current absence of any clear hints for new physics, it is clear that they cannot be used to judge the likelihood of experimental discoveries, as there is no objective way to evaluate these criteria.  
Both criteria strongly depend on the choice of parameterization, the first one in addition relies on a subjective definition of what is natural. 
It is nevertheless instructive to apply them to the parameter space under consideration. 
Common choices for a notion of ``naturalness'' are  $a)$ the idea that all parameters should be of the same order 
or $b)$ the stability against radiative corrections \cite{'tHooft:1979bh}.\footnote{Though well-though theoretical arguments for these choices can be made, it should be kept in mind that $a)$ is not fulfilled in the SM in its standard parameterizations and $b)$ implicitly assumes that nature makes a qualitative difference between what we call the \emph{classical theory} and \emph{quantum corrections}, being somewhat at odds with the idea that only observable quantities are meaningful in quantum mechanics. }

By virtue of the smallness of the Yukawa interactions, the model considered here is natural in the sense of $b)$ in the region below the red line in FIG.~\ref{bounds}, cf. Eq.(\ref{RadiativeCorrections}). 
The mass spectrum is natural in the sense of $a)$, as all masses and mass differences are roughly of the same order in most of the parameters pace we studied. 
To apply criterion $a)$ to the Yukawa coupling matrix $F$ we will use two different approaches, in order to at least partly alleviate the parameterization dependence.

We first compare the quantity $y\equiv\sqrt{{\rm tr}(F^\dagger F)}$ to the "naive seesaw expectation" $y_0=\sqrt{m_{\rm atm}M_M/v^2}$, i.e. the would-be value of the Yukawa coupling in a world without flavour, $n=1$ and a neutrino mass of the order $m_{\rm atm}$. 
We consider the representative mass choice $M_1=1$ GeV, $M_2=2$ GeV, $M_3=3$ GeV. For a typical point near the current limit from BELLE of $|\Theta_{\mu2}|^2< 3\times 10^{-5}$ we get $y\simeq 5\times 10^{-5}$
For a Majorana mass of $2$ GeV the naive seesaw expectation would yield $y_0= (2{\rm GeV} \times m_{\rm atm}/v^2)^{1/2}\simeq 5.7\times 10^{-8}$. 
The ratio $y/y_0\simeq1:845$ of these two values implies that there are cancellations in the eigenvalues of the matrix $m_\nu m_\nu^\dagger$ in this parameterization, which keep the neutrino masses small in spite of relatively large couplings of the $N_I$. 
For points near the anticipated SHiP sensitivity $|\Theta_{\mu2}|^2\sim 10^{-9}$, with a mass $M_2=1.5$ GeV in the regime where this sensitivity can be achieved, we find $y\simeq 1.2\times 10^{-6}$, a value roughly 25 times larger than $y_0$. 
It was suggested in \cite{Shuve:2014zua} that the square of this ratio should be taken as a measure to quantify these cancellations, since the physical neutrino masses scale as $\propto y^2$. This gives ratios of roughly $(y/y_0)^2\simeq 7.1\times10^5$ and 
$(y/y_0)^2\simeq 625$, respectively.
One possible interpretation for the relatively large deviations from $y_0$ is that leptogenesis can only occur in a ``special'' region in parameter space. At this stage it is not obvious whether this region can be distinct from others by some approximate symmetry or some other consideration. Another possible interpretation is that the cancellations of large numbers in the neutrino mass matrix simply indicates that we chose an inconvenient parameterization.

In a second approach we directly consider the individual parameter values. The most important parameters for both, the CP-violation and the active-sterile mixing, are the imaginary parts of the $\omega_{ij}$.
In principle these can take values from $-\infty$ to $\infty$, but for values of $|{\rm Im}\omega_{ij}|$ much bigger than the interval we consider radiative corrections to the neutrino masses become large.
Typical values that lead to leptogenesis for a $|\Theta_{\mu 2}|^2$ near the BELLE limit are $|{\rm Im}\omega_{12}|\sim |{\rm Im}\omega_{13}|\sim 4$ and $|{\rm Im}\omega_{23}|\lesssim 1$, i.e. are all roughly of the same order, but show a clear pattern.
This reflects the requirement that the electron flavour should be protected from washout, and the pattern may be different if we had chosen another hierarchy in equations (\ref{electronsmall}) and (\ref{assumption}). 
In the region of smaller $|\Theta_{\mu 2}|^2$ near the SHiP sensitivity, the $|{\rm Im}\omega_{ij}|$ 
do not show such a clear pattern and are generally of order one. This can easily be explained, as in this region the $N_I$-equilibration only occurs shortly before sphaleron freezeout and the criterion of an incomplete washout is fulfilled for a wider choice of parameters. 

We interpret our results as an indication that there are considerable regions in the parameter space we investigated in which leptogenesis is possible. 
How extended these are strongly depends on the measure used in the parameter space. In lack of any knowledge about new physics beyond (\ref{L}) that might govern the values of the masses and couplings, we take a bottom-up approach that is agnostic with respect to parameterization and express our results in terms of the observable active-sterile mixing in FIG.~\ref{bounds}.
As far as observables are concerned, an interesting parameter region for $M_2$ in regime $ii)$ is the one with $|\Theta_{\mu I}|^2\sim 10^{-5}$ near the experimental bounds. Here leptogenesis is possible when the $U_e^2\equiv\sum_I|\Theta_{e I}|^2$ is considerably smaller than $U_\mu^2$, roughly by four orders of magnitude. 
For values of $|\Theta_{\mu I}|^2$ near the anticipated SHiP sensitivity this difference reduces to a factor of order one. It could be that even $|\Theta_{\mu I}|^2<|\Theta_{e I}|^2$ is allowed in this region, but our assumption (\ref{electronsmall}) prevents us from exploring this possibility within the current approach.
The region with small $|\Theta_{e I}|^2$ is cosmologically motivated because it leads to a flavour-asymmetric washout of the lepton asymmetries in the early universe, which allows to have observably large mixings with other flavours. At the same time it automatically implies consistency with the bounds on (\ref{BranchingRatio}) and (\ref{mee}).
It would be interesting to see if this region is characterized by a symmetry, such as an approximately conserved lepton number that is responsible for the smallness of neutrino masses, or some other model building consideration.
We postpone the investigation of this question to future work.

\section{Conclusions}\label{sec:conclusions}
We have shown that three RH neutrinos in the type-I seesaw model (\ref{L}) with masses in the GeV range and experimentally accessible mixings can explain the BAU via leptogenesis.
No degeneracy in the Majorana masses is required. In the limit of degenerate Majorana masses the model is expected to resemble many properties of the well-known $\nu$MSM. 
For a non-degenerate mass spectrum makes this scenario is clearly distinguishable from the $\nu$MSM and other realizations of resonant leptogenesis.

A discovery of heavy neutral leptons at LHCb or BELLE would be smoking gun evidence that these particles can be the common origin of matter in the universe and the observed neutrino masses.
Both of these experiments have already entered the cosmologically interesting parameter space and will continue to take data after the updates currently under way. 
The chances for a discovery can be optimized by studying all possible decay channels of $B$-mesons that involve $N_I$.
The perspectives would be even better at the proposed SHiP experiment, for which our findings provide strong motivation.

\section*{Acknowledgements}
We are grateful to Nicola Serra, Walter Bonivento, Artem Ivashko, Phillip Urquijo, Dmitri Liventsev and Junji Hisano for experimental data and explanations.  
We would also like to thank Mikhail Shaposhnikov and Dmitry Gorbunov for helpful discussions. 
MaD thanks CERN for the hospitality during the final stage of this project.
This work is supported by the Gottfried Wilhelm Leibniz programme of the Deutsche Forschungsgemeinschaft (DFG) and by the DFG cluster of excellence Origin and Structure of the Universe.

\bibliographystyle{JHEP}
\bibliography{all}

\providecommand{\href}[2]{#2}\begingroup\raggedright\begin{thebibliography}{10%
0}

\bibitem{Abazajian:2012ys}
K.~Abazajian, M.~Acero, S.~Agarwalla, A.~Aguilar-Arevalo, C.~Albright, et~al.,
  {\it {Light Sterile Neutrinos: A White Paper}},
  \href{http://arxiv.org/abs/1204.5379}{{\tt arXiv:1204.5379}}.

\bibitem{Drewes:2013gca}
M.~Drewes, {\it {The Phenomenology of Right Handed Neutrinos}},  {\em
  International Journal of Modern Physics E, Vol.} {\bf 22} (2013) 1330019,
  [\href{http://arxiv.org/abs/1303.6912}{{\tt arXiv:1303.6912}}].

\bibitem{Minkowski:1977sc}
P.~Minkowski, {\it mu $\to$ e gamma at a rate of one out of 1-billion muon
  decays?},  {\em Phys. Lett.} {\bf B67} (1977) 421.

\bibitem{GellMann:seesaw}
M.~Gell-Mann, P.~Ramond, and R.~Slansky {\em in Supergravity, ed. by D.
  Freedman et al., North Holland} (1979).

\bibitem{Mohapatra:1979ia}
R.~N. Mohapatra and G.~Senjanovic, {\it {Neutrino mass and spontaneous parity
  nonconservation}},  {\em Phys. Rev. Lett.} {\bf 44} (1980) 912.

\bibitem{Yanagida:1980xy}
T.~Yanagida, {\it {Horizontal Symmetry and Masses of Neutrinos}},  {\em
  Prog.Theor.Phys.} {\bf 64} (1980) 1103.

\bibitem{Fukugita:1986hr}
M.~Fukugita and T.~Yanagida, {\it {Baryogenesis Without Grand Unification}},
  {\em Phys. Lett.} {\bf B174} (1986) 45.

\bibitem{Davidson:2002qv}
S.~Davidson and A.~Ibarra, {\it {A Lower bound on the right-handed neutrino
  mass from leptogenesis}},  {\em Phys.Lett.} {\bf B535} (2002) 25--32,
  [\href{http://arxiv.org/abs/hep-ph/0202239}{{\tt hep-ph/0202239}}].

\bibitem{Antusch:2009gn}
S.~Antusch, S.~Blanchet, M.~Blennow, and E.~Fernandez-Martinez, {\it
  {Non-unitary Leptonic Mixing and Leptogenesis}},  {\em JHEP} {\bf 1001}
  (2010) 017, [\href{http://arxiv.org/abs/0910.5957}{{\tt arXiv:0910.5957}}].

\bibitem{Racker:2012vw}
J.~Racker, M.~Pena, and N.~Rius, {\it {Leptogenesis with small violation of
  B-L}},  {\em JCAP} {\bf 1207} (2012) 030,
  [\href{http://arxiv.org/abs/1205.1948}{{\tt arXiv:1205.1948}}].

\bibitem{Asaka:2005an}
T.~Asaka, S.~Blanchet, and M.~Shaposhnikov, {\it The $\nu$msm, dark matter and
  neutrino masses},  {\em Phys. Lett.} {\bf B631} (2005) 151--156,
  [\href{http://arxiv.org/abs/hep-ph/0503065}{{\tt hep-ph/0503065}}].

\bibitem{Asaka:2005pn}
T.~Asaka and M.~Shaposhnikov, {\it The $\nu$msm, dark matter and baryon
  asymmetry of the universe},  {\em Phys. Lett.} {\bf B620} (2005) 17--26,
  [\href{http://arxiv.org/abs/hep-ph/0505013}{{\tt hep-ph/0505013}}].

\bibitem{Boyarsky:2009ix}
A.~Boyarsky, O.~Ruchayskiy, and M.~Shaposhnikov, {\it {The role of sterile
  neutrinos in cosmology and astrophysics}},  {\em Ann. Rev. Nucl. Part. Sci.}
  {\bf 59} (2009) 191--214, [\href{http://arxiv.org/abs/0901.0011}{{\tt
  arXiv:0901.0011}}].

\bibitem{Canetti:2012kh}
L.~Canetti, M.~Drewes, T.~Frossard, and M.~Shaposhnikov, {\it {Dark Matter,
  Baryogenesis and Neutrino Oscillations from Right Handed Neutrinos}},  {\em
  Phys.Rev.} {\bf D87} (2013), no.~9 093006,
  [\href{http://arxiv.org/abs/1208.4607}{{\tt arXiv:1208.4607}}].

\bibitem{Pilaftsis:1997jf}
A.~Pilaftsis, {\it {CP violation and baryogenesis due to heavy Majorana
  neutrinos}},  {\em Phys.Rev.} {\bf D56} (1997) 5431--5451,
  [\href{http://arxiv.org/abs/hep-ph/9707235}{{\tt hep-ph/9707235}}].

\bibitem{Pilaftsis:2003gt}
A.~Pilaftsis and T.~E. Underwood, {\it {Resonant leptogenesis}},  {\em
  Nucl.Phys.} {\bf B692} (2004) 303--345,
  [\href{http://arxiv.org/abs/hep-ph/0309342}{{\tt hep-ph/0309342}}].

\bibitem{Canetti:2010aw}
L.~Canetti and M.~Shaposhnikov, {\it {Baryon Asymmetry of the Universe in the
  NuMSM}},  {\em JCAP} {\bf 1009} (2010) 001,
  [\href{http://arxiv.org/abs/1006.0133}{{\tt arXiv:1006.0133}}].

\bibitem{Asaka:2011wq}
T.~Asaka, S.~Eijima, and H.~Ishida, {\it {Kinetic Equations for Baryogenesis
  via Sterile Neutrino Oscillation}},  {\em JCAP} {\bf 1202} (2012) 021,
  [\href{http://arxiv.org/abs/1112.5565}{{\tt arXiv:1112.5565}}].

\bibitem{Canetti:2012vf}
L.~Canetti, M.~Drewes, and M.~Shaposhnikov, {\it {Sterile Neutrinos as the
  Origin of Dark and Baryonic Matter}},  {\em Phys. Rev. Lett.} {\bf 110}
  (2013) 061801, [\href{http://arxiv.org/abs/1204.3902}{{\tt
  arXiv:1204.3902}}].

\bibitem{Shuve:2014zua}
B.~Shuve and I.~Yavin, {\it {Baryogenesis through Neutrino Oscillations: A
  Unified Perspective}},  {\em Phys.Rev.} {\bf D89} (2014) 075014,
  [\href{http://arxiv.org/abs/1401.2459}{{\tt arXiv:1401.2459}}].

\bibitem{Garbrecht:2014bfa}
B.~Garbrecht, {\it {More Viable Parameter Space for Leptogenesis}},
  \href{http://arxiv.org/abs/1401.3278}{{\tt arXiv:1401.3278}}.

\bibitem{Kersten:2007vk}
J.~Kersten and A.~Y. Smirnov, {\it {Right-Handed Neutrinos at CERN LHC and the
  Mechanism of Neutrino Mass Generation}},  {\em Phys.Rev.} {\bf D76} (2007)
  073005, [\href{http://arxiv.org/abs/0705.3221}{{\tt arXiv:0705.3221}}].

\bibitem{Ibarra:2011xn}
A.~Ibarra, E.~Molinaro, and S.~Petcov, {\it {Low Energy Signatures of the TeV
  Scale See-Saw Mechanism}},  {\em Phys.Rev.} {\bf D84} (2011) 013005,
  [\href{http://arxiv.org/abs/1103.6217}{{\tt arXiv:1103.6217}}].

\bibitem{Bonivento:2013jag}
W.~Bonivento, A.~Boyarsky, H.~Dijkstra, U.~Egede, M.~Ferro-Luzzi, et~al., {\it
  {Proposal to Search for Heavy Neutral Leptons at the SPS}},
  \href{http://arxiv.org/abs/1310.1762}{{\tt arXiv:1310.1762}}.

\bibitem{deGouvea:2005er}
A.~de~Gouvea, {\it {See-saw energy scale and the LSND anomaly}},  {\em
  Phys.Rev.} {\bf D72} (2005) 033005,
  [\href{http://arxiv.org/abs/hep-ph/0501039}{{\tt hep-ph/0501039}}].

\bibitem{Casas:2001sr}
J.~Casas and A.~Ibarra, {\it {Oscillating neutrinos and muon to e, gamma}},
  {\em Nucl.Phys.} {\bf B618} (2001) 171--204,
  [\href{http://arxiv.org/abs/hep-ph/0103065}{{\tt hep-ph/0103065}}].

\bibitem{Sakharov:1967dj}
A.~D. Sakharov, {\it {Violation of CP Invariance, c Asymmetry, and Baryon
  Asymmetry of the Universe}},  {\em Pisma Zh. Eksp. Teor. Fiz.} {\bf 5} (1967)
  32--35.

\bibitem{Akhmedov:1998qx}
E.~K. Akhmedov, V.~A. Rubakov, and A.~Y. Smirnov, {\it Baryogenesis via
  neutrino oscillations},  {\em Phys. Rev. Lett.} {\bf 81} (1998) 1359--1362,
  [\href{http://arxiv.org/abs/hep-ph/9803255}{{\tt hep-ph/9803255}}].

\bibitem{Bezrukov:2008ut}
F.~Bezrukov, D.~Gorbunov, and M.~Shaposhnikov, {\it {On initial conditions for
  the Hot Big Bang}},  {\em JCAP} {\bf 0906} (2009) 029,
  [\href{http://arxiv.org/abs/0812.3622}{{\tt arXiv:0812.3622}}].

\bibitem{Kuzmin:1985mm}
V.~A. Kuzmin, V.~A. Rubakov, and M.~E. Shaposhnikov, {\it On the anomalous
  electroweak baryon number nonconservation in the early universe},  {\em Phys.
  Lett.} {\bf B155} (1985) 36.

\bibitem{Burnier:2005hp}
Y.~Burnier, M.~Laine, and M.~Shaposhnikov, {\it {Baryon and lepton number
  violation rates across the electroweak crossover}},  {\em JCAP} {\bf 0602}
  (2006) 007, [\href{http://arxiv.org/abs/hep-ph/0511246}{{\tt
  hep-ph/0511246}}].

\bibitem{D'Onofrio:2012jk}
M.~D'Onofrio, K.~Rummukainen, and A.~Tranberg, {\it {The Sphaleron Rate through
  the Electroweak Cross-over}},  {\em JHEP} {\bf 1208} (2012) 123,
  [\href{http://arxiv.org/abs/1207.0685}{{\tt arXiv:1207.0685}}].

\bibitem{D'Onofrio:2014kta}
M.~D'Onofrio, K.~Rummukainen, and A.~Tranberg, {\it {The Sphaleron Rate in the
  Minimal Standard Model}},  \href{http://arxiv.org/abs/1404.3565}{{\tt
  arXiv:1404.3565}}.

\bibitem{Drewes:2012ma}
M.~Drewes and B.~Garbrecht, {\it {Leptogenesis from a GeV Seesaw without Mass
  Degeneracy}},  {\em JHEP} {\bf 1303} (2013) 096,
  [\href{http://arxiv.org/abs/1206.5537}{{\tt arXiv:1206.5537}}].

\bibitem{Khoze:2013oga}
V.~V. Khoze and G.~Ro, {\it {Leptogenesis and Neutrino Oscillations in the
  Classically Conformal Standard Model with the Higgs Portal}},  {\em JHEP}
  {\bf 1310} (2013) 075, [\href{http://arxiv.org/abs/1307.3764}{{\tt
  arXiv:1307.3764}}].

\bibitem{Gorbunov:2007ak}
D.~Gorbunov and M.~Shaposhnikov, {\it {How to find neutral leptons of the
  $\nu$MSM?}},  {\em JHEP} {\bf 10} (2007) 015,
  [\href{http://arxiv.org/abs/0705.1729}{{\tt arXiv:0705.1729}}].

\bibitem{Shaposhnikov:2008pf}
M.~Shaposhnikov, {\it {The $\nu$MSM, leptonic asymmetries, and properties of
  singlet fermions}},  {\em JHEP} {\bf 08} (2008) 008,
  [\href{http://arxiv.org/abs/0804.4542}{{\tt arXiv:0804.4542}}].

\bibitem{Asaka:2011pb}
T.~Asaka, S.~Eijima, and H.~Ishida, {\it {Mixing of Active and Sterile
  Neutrinos}},  {\em JHEP} {\bf 1104} (2011) 011,
  [\href{http://arxiv.org/abs/1101.1382}{{\tt arXiv:1101.1382}}].

\bibitem{Gorbunov:2013dta}
D.~Gorbunov and A.~Panin, {\it {On the minimal active-sterile neutrino mixing
  in seesaw type I mechanism with sterile neutrinos at GeV scale}},  {\em
  Phys.Rev.} {\bf D89} (2014) 017302,
  [\href{http://arxiv.org/abs/1312.2887}{{\tt arXiv:1312.2887}}].

\bibitem{Garny:2011hg}
M.~Garny, A.~Kartavtsev, and A.~Hohenegger, {\it {Leptogenesis from first
  principles in the resonant regime}},
  \href{http://arxiv.org/abs/1112.6428}{{\tt arXiv:1112.6428}}.

\bibitem{Garbrecht:2011aw}
B.~Garbrecht and M.~Herranen, {\it {Effective Theory of Resonant Leptogenesis
  in the Closed-Time-Path Approach}},  {\em Nucl.Phys.} {\bf B861} (2012)
  17--52, [\href{http://arxiv.org/abs/1112.5954}{{\tt arXiv:1112.5954}}].

\bibitem{Iso:2013lba}
S.~Iso, K.~Shimada, and M.~Yamanaka, {\it {Kadanoff-Baym approach to the
  thermal resonant leptogenesis}},  \href{http://arxiv.org/abs/1312.7680}{{\tt
  arXiv:1312.7680}}.

\bibitem{Dev:2014laa}
P.~S.~B. Dev, P.~Millington, A.~Pilaftsis, and D.~Teresi, {\it {Flavour
  Covariant Transport Equations: an Application to Resonant Leptogenesis}},
  \href{http://arxiv.org/abs/1404.1003}{{\tt arXiv:1404.1003}}.

\bibitem{Iso:2014afa}
S.~Iso and K.~Shimada, {\it {Coherent Flavour Oscillation and CP Violating
  Parameter in Thermal Resonant Leptogenesis}},
  \href{http://arxiv.org/abs/1404.4816}{{\tt arXiv:1404.4816}}.

\bibitem{Hohenegger:2014cpa}
A.~Hohenegger and A.~Kartavtsev, {\it {Leptogenesis in crossing and runaway
  regimes}},  \href{http://arxiv.org/abs/1404.5309}{{\tt arXiv:1404.5309}}.

\bibitem{Buchmuller:2000nd}
W.~Buchmuller and S.~Fredenhagen, {\it {Quantum mechanics of baryogenesis}},
  {\em Phys.Lett.} {\bf B483} (2000) 217--224,
  [\href{http://arxiv.org/abs/hep-ph/0004145}{{\tt hep-ph/0004145}}].

\bibitem{DeSimone:2007rw}
A.~De~Simone and A.~Riotto, {\it {Quantum Boltzmann Equations and
  Leptogenesis}},  {\em JCAP} {\bf 0708} (2007) 002,
  [\href{http://arxiv.org/abs/hep-ph/0703175}{{\tt hep-ph/0703175}}].

\bibitem{Garny:2009rv}
M.~Garny, A.~Hohenegger, A.~Kartavtsev, and M.~Lindner, {\it {Systematic
  approach to leptogenesis in nonequilibrium QFT: Vertex contribution to the
  CP-violating parameter}},  {\em Phys.Rev.} {\bf D80} (2009) 125027,
  [\href{http://arxiv.org/abs/0909.1559}{{\tt arXiv:0909.1559}}].

\bibitem{Garny:2009qn}
M.~Garny, A.~Hohenegger, A.~Kartavtsev, and M.~Lindner, {\it {Systematic
  approach to leptogenesis in nonequilibrium QFT: Self-energy contribution to
  the CP-violating parameter}},  {\em Phys.Rev.} {\bf D81} (2010) 085027,
  [\href{http://arxiv.org/abs/0911.4122}{{\tt arXiv:0911.4122}}].

\bibitem{Anisimov:2010aq}
A.~Anisimov, W.~Buchmuller, M.~Drewes, and S.~Mendizabal, {\it {Leptogenesis
  from Quantum Interference in a Thermal Bath}},  {\em Phys.Rev.Lett.} {\bf
  104} (2010) 121102, [\href{http://arxiv.org/abs/1001.3856}{{\tt
  arXiv:1001.3856}}].

\bibitem{Garny:2010nj}
M.~Garny, A.~Hohenegger, and A.~Kartavtsev, {\it {Medium corrections to the
  CP-violating parameter in leptogenesis}},  {\em Phys.Rev.} {\bf D81} (2010)
  085028, [\href{http://arxiv.org/abs/1002.0331}{{\tt arXiv:1002.0331}}].

\bibitem{Beneke:2010wd}
M.~Beneke, B.~Garbrecht, M.~Herranen, and P.~Schwaller, {\it {Finite Number
  Density Corrections to Leptogenesis}},  {\em Nucl.Phys.} {\bf B838} (2010)
  1--27, [\href{http://arxiv.org/abs/1002.1326}{{\tt arXiv:1002.1326}}].

\bibitem{Garny:2010nz}
M.~Garny, A.~Hohenegger, and A.~Kartavtsev, {\it {Quantum corrections to
  leptogenesis from the gradient expansion}},
  \href{http://arxiv.org/abs/1005.5385}{{\tt arXiv:1005.5385}}.

\bibitem{Garbrecht:2010sz}
B.~Garbrecht, {\it {Leptogenesis: The Other Cuts}},  {\em Nucl.Phys.} {\bf
  B847} (2011) 350--366, [\href{http://arxiv.org/abs/1011.3122}{{\tt
  arXiv:1011.3122}}].

\bibitem{Beneke:2010dz}
M.~Beneke, B.~Garbrecht, C.~Fidler, M.~Herranen, and P.~Schwaller, {\it
  {Flavoured Leptogenesis in the CTP Formalism}},  {\em Nucl.Phys.} {\bf B843}
  (2011) 177--212, [\href{http://arxiv.org/abs/1007.4783}{{\tt
  arXiv:1007.4783}}].

\bibitem{Anisimov:2010dk}
A.~Anisimov, W.~Buchmuller, M.~Drewes, and S.~Mendizabal, {\it {Quantum
  Leptogenesis I}},  {\em Annals Phys.} {\bf 326} (2011) 1998--2038,
  [\href{http://arxiv.org/abs/1012.5821}{{\tt arXiv:1012.5821}}].

\bibitem{Fidler:2011yq}
C.~Fidler, M.~Herranen, K.~Kainulainen, and P.~M. Rahkila, {\it {Flavoured
  quantum Boltzmann equations from cQPA}},  {\em JHEP} {\bf 1202} (2012) 065,
  [\href{http://arxiv.org/abs/1108.2309}{{\tt arXiv:1108.2309}}].

\bibitem{Garbrecht:2012qv}
B.~Garbrecht, {\it {Leptogenesis from Additional Higgs Doublets}},  {\em
  Phys.Rev.} {\bf D85} (2012) 123509,
  [\href{http://arxiv.org/abs/1201.5126}{{\tt arXiv:1201.5126}}].

\bibitem{Frossard:2012pc}
T.~Frossard, M.~Garny, A.~Hohenegger, A.~Kartavtsev, and D.~Mitrouskas, {\it
  {Systematic approach to thermal leptogenesis}},  {\em Phys.Rev.} {\bf D87}
  (2013) 085009, [\href{http://arxiv.org/abs/1211.2140}{{\tt
  arXiv:1211.2140}}].

\bibitem{Sigl:1992fn}
G.~Sigl and G.~Raffelt, {\it General kinetic description of relativistic mixed
  neutrinos},  {\em Nucl. Phys.} {\bf B406} (1993) 423--451.

\bibitem{Khlebnikov:1988sr}
S.~Y. Khlebnikov and M.~Shaposhnikov, {\it {The Statistical Theory of Anomalous
  Fermion Number Nonconservation}},  {\em Nucl.Phys.} {\bf B308} (1988)
  885--912.

\bibitem{Laine:1999wv}
M.~Laine and M.~E. Shaposhnikov, {\it {A Remark on sphaleron erasure of baryon
  asymmetry}},  {\em Phys.Rev.} {\bf D61} (2000) 117302,
  [\href{http://arxiv.org/abs/hep-ph/9911473}{{\tt hep-ph/9911473}}].

\bibitem{Capozzi:2013csa}
F.~Capozzi, G.~Fogli, E.~Lisi, A.~Marrone, D.~Montanino, et~al., {\it {Status
  of three-neutrino oscillation parameters, circa 2013}},
  \href{http://arxiv.org/abs/1312.2878}{{\tt arXiv:1312.2878}}.

\bibitem{AristizabalSierra:2011mn}
D.~Aristizabal~Sierra and C.~E. Yaguna, {\it {On the importance of the 1-loop
  finite corrections to seesaw neutrino masses}},  {\em JHEP} {\bf 1108} (2011)
  013, [\href{http://arxiv.org/abs/1106.3587}{{\tt arXiv:1106.3587}}].

\bibitem{Pilaftsis:1991ug}
A.~Pilaftsis, {\it {Radiatively induced neutrino masses and large Higgs
  neutrino couplings in the standard model with Majorana fields}},  {\em
  Z.Phys.} {\bf C55} (1992) 275--282,
  [\href{http://arxiv.org/abs/hep-ph/9901206}{{\tt hep-ph/9901206}}].

\bibitem{Besak:2012qm}
D.~Besak and D.~Bodeker, {\it {Thermal production of ultrarelativistic
  right-handed neutrinos: Complete leading-order results}},  {\em JCAP} {\bf
  1203} (2012) 029, [\href{http://arxiv.org/abs/1202.1288}{{\tt
  arXiv:1202.1288}}].

\bibitem{Garbrecht:2013bia}
B.~Garbrecht, F.~Glowna, and P.~Schwaller, {\it {Scattering Rates For
  Leptogenesis: Damping of Lepton Flavour Coherence and Production of Singlet
  Neutrinos}},  \href{http://arxiv.org/abs/1303.5498}{{\tt arXiv:1303.5498}}.

\bibitem{Bodeker:2014hqa}
D.~Bodeker and M.~Laine, {\it {Kubo relations and radiative corrections for
  lepton number washout}},  \href{http://arxiv.org/abs/1403.2755}{{\tt
  arXiv:1403.2755}}.

\bibitem{Laine:2013lka}
M.~Laine, {\it {Thermal right-handed neutrino production rate in the
  relativistic regime}},  {\em JHEP} {\bf 1308} (2013) 138,
  [\href{http://arxiv.org/abs/1307.4909}{{\tt arXiv:1307.4909}}].

\bibitem{Canetti:2012zc}
L.~Canetti, M.~Drewes, and M.~Shaposhnikov, {\it {Matter and Antimatter in the
  Universe}},  {\em New J. Phys.} {\bf 14} (2012) 095012,
  [\href{http://arxiv.org/abs/1204.4186}{{\tt arXiv:1204.4186}}].

\bibitem{Ade:2013zuv}
{\bf Planck Collaboration} Collaboration, P.~Ade et~al., {\it {Planck 2013
  results. XVI. Cosmological parameters}},
  \href{http://arxiv.org/abs/1303.5076}{{\tt arXiv:1303.5076}}.

\bibitem{Canetti:2013qna}
L.~Canetti and M.~Shaposhnikov, {\it {The $\nu$ MSM and muon to electron
  conversion experiments}},  {\em Hyperfine Interact.} {\bf 214} (2013),
  no.~1-3 5--11.

\bibitem{Adam:2013mnn}
{\bf MEG Collaboration} Collaboration, J.~Adam et~al., {\it {New constraint on
  the existence of the $\mu^+ \to e^+\gamma$ decay}},  {\em Phys.Rev.Lett.}
  {\bf 110} (2013), no.~20 201801, [\href{http://arxiv.org/abs/1303.0754}{{\tt
  arXiv:1303.0754}}].

\bibitem{Bezrukov:2005mx}
F.~Bezrukov, {\it {nu MSM-predictions for neutrinoless double beta decay}},
  {\em Phys.Rev.} {\bf D72} (2005) 071303,
  [\href{http://arxiv.org/abs/hep-ph/0505247}{{\tt hep-ph/0505247}}].

\bibitem{LopezPavon:2012zg}
J.~Lopez-Pavon, S.~Pascoli, and C.-f. Wong, {\it {Can heavy neutrinos dominate
  neutrinoless double beta decay?}},
  \href{http://arxiv.org/abs/1209.5342}{{\tt arXiv:1209.5342}}.

\bibitem{Asaka:2013jfa}
T.~Asaka and S.~Eijima, {\it {Direct Search for Right-handed Neutrinos and
  Neutrinoless Double Beta Decay}},  \href{http://arxiv.org/abs/1308.3550}{{\tt
  arXiv:1308.3550}}.

\bibitem{Merle:2013ibc}
A.~Merle and V.~Niro, {\it {Influence of a keV sterile neutrino on
  neutrino-less double beta decay -- how things changed in the recent years}},
  \href{http://arxiv.org/abs/1302.2032}{{\tt arXiv:1302.2032}}.

\bibitem{Girardi:2013zra}
I.~Girardi, A.~Meroni, and S.~Petcov, {\it {Neutrinoless Double Beta Decay in
  the Presence of Light Sterile Neutrinos}},
  \href{http://arxiv.org/abs/1308.5802}{{\tt arXiv:1308.5802}}.

\bibitem{Agostini:2013mzu}
{\bf GERDA Collaboration} Collaboration, M.~Agostini et~al., {\it {Results on
  Neutrinoless Double-$\beta$ Decay of $^{76}$Ge from Phase I of the GERDA
  Experiment}},  {\em Phys.Rev.Lett.} {\bf 111} (2013), no.~12 122503,
  [\href{http://arxiv.org/abs/1307.4720}{{\tt arXiv:1307.4720}}].

\bibitem{Dinh:2012bp}
D.~Dinh, A.~Ibarra, E.~Molinaro, and S.~Petcov, {\it {The $\mu - e$ Conversion
  in Nuclei, $\mu \to e \gamma, \mu \to 3e$ Decays and TeV Scale See-Saw
  Scenarios of Neutrino Mass Generation}},  {\em JHEP} {\bf 1208} (2012) 125,
  [\href{http://arxiv.org/abs/1205.4671}{{\tt arXiv:1205.4671}}].

\bibitem{Lello:2012in}
L.~Lello and D.~Boyanovsky, {\it {Charged lepton mixing via heavy sterile
  neutrinos}},  {\em Nucl.Phys.} {\bf B880} (2014) 109--133,
  [\href{http://arxiv.org/abs/1212.4167}{{\tt arXiv:1212.4167}}].

\bibitem{Cheng:1980tp}
T.~Cheng and L.-F. Li, {\it {$\mu \rightarrow e \gamma$ IN THEORIES WITH DIRAC
  AND MAJORANA NEUTRINO MASS TERMS}},  {\em Phys.Rev.Lett.} {\bf 45} (1980)
  1908.

\bibitem{Bilenky:1977du}
S.~M. Bilenky, S.~Petcov, and B.~Pontecorvo, {\it {Lepton Mixing, $\mu
  \rightarrow e + \gamma$ Decay and Neutrino Oscillations}},  {\em Phys.Lett.}
  {\bf B67} (1977) 309.

\bibitem{Bernardi:1987ek}
G.~Bernardi et~al., {\it Further limits on heavy neutrino couplings},  {\em
  Phys. Lett.} {\bf B203} (1988) 332.

\bibitem{Vaitaitis:1999wq}
{\bf NuTeV} Collaboration, A.~Vaitaitis et~al., {\it {Search for neutral heavy
  leptons in a high-energy neutrino beam}},  {\em Phys. Rev. Lett.} {\bf 83}
  (1999) 4943--4946, [\href{http://arxiv.org/abs/hep-ex/9908011}{{\tt
  hep-ex/9908011}}].

\bibitem{Ruchayskiy:2011aa}
O.~Ruchayskiy and A.~Ivashko, {\it {Experimental bounds on sterile neutrino
  mixing angles}},  {\em JHEP} {\bf 1206} (2012) 100,
  [\href{http://arxiv.org/abs/1112.3319}{{\tt arXiv:1112.3319}}].

\bibitem{Badier:1985wg}
{\bf NA3 Collaboration} Collaboration, J.~Badier et~al., {\it {Direct Photon
  Production From Pions and Protons at 200-{GeV}/$c$}},  {\em Z.Phys.} {\bf
  C31} (1986) 341.

\bibitem{Vilain:1994vg}
{\bf CHARM II Collaboration} Collaboration, P.~Vilain et~al., {\it {Search for
  heavy isosinglet neutrinos}},  {\em Phys.Lett.} {\bf B343} (1995) 453--458.

\bibitem{Abreu:1996pa}
{\bf DELPHI Collaboration} Collaboration, P.~Abreu et~al., {\it {Search for
  neutral heavy leptons produced in Z decays}},  {\em Z.Phys.} {\bf C74} (1997)
  57--71.

\bibitem{Atre:2009rg}
A.~Atre, T.~Han, S.~Pascoli, and B.~Zhang, {\it {The Search for Heavy Majorana
  Neutrinos}},  {\em JHEP} {\bf 0905} (2009) 030,
  [\href{http://arxiv.org/abs/0901.3589}{{\tt arXiv:0901.3589}}].

\bibitem{Abada:2012mc}
A.~Abada, D.~Das, A.~Teixeira, A.~Vicente, and C.~Weiland, {\it {Tree-level
  lepton universality violation in the presence of sterile neutrinos: impact
  for $R_K$ and $R_\pi$}},  {\em JHEP} {\bf 1302} (2013) 048,
  [\href{http://arxiv.org/abs/1211.3052}{{\tt arXiv:1211.3052}}].

\bibitem{Abada:2013aba}
A.~Abada, A.~Teixeira, A.~Vicente, and C.~Weiland, {\it {Sterile neutrinos in
  leptonic and semileptonic decays}},  {\em JHEP} {\bf 1402} (2014) 091,
  [\href{http://arxiv.org/abs/1311.2830}{{\tt arXiv:1311.2830}}].

\bibitem{Basso:2013jka}
L.~Basso, O.~Fischer, and J.~J. van~der Bij, {\it {Precision tests of unitarity
  in leptonic mixing}},  {\em Europhys.Lett.} {\bf 105} (2014) 11001,
  [\href{http://arxiv.org/abs/1310.2057}{{\tt arXiv:1310.2057}}].

\bibitem{Endo:2014hza}
M.~Endo and T.~Yoshinaga, {\it {Lepton Universality Test of Extra Leptons using
  Hadron Decay}},  \href{http://arxiv.org/abs/1404.4498}{{\tt
  arXiv:1404.4498}}.

\bibitem{Aaij:2014aba}
{\bf LHCb collaboration} Collaboration, R.~Aaij et~al., {\it {Search for
  Majorana neutrinos in $B^- \to \pi^+\mu^-\mu^-$ decays}},  {\em
  Phys.Rev.Lett.} {\bf 112} (2014) 131802,
  [\href{http://arxiv.org/abs/1401.5361}{{\tt arXiv:1401.5361}}].

\bibitem{Liventsev:2013zz}
{\bf Belle Collaboration} Collaboration, D.~Liventsev et~al., {\it {Search for
  heavy neutrinos at Belle}},  {\em Phys.Rev.} {\bf D87} (2013), no.~7 071102,
  [\href{http://arxiv.org/abs/1301.1105}{{\tt arXiv:1301.1105}}].

\bibitem{Asaka:2012bb}
T.~Asaka, S.~Eijima, and A.~Watanabe, {\it {Heavy neutrino search in
  accelerator-based experiments}},  \href{http://arxiv.org/abs/1212.1062}{{\tt
  arXiv:1212.1062}}.

\bibitem{Alekhin:2015oba}
S.~Alekhin, W.~Altmannshofer, T.~Asaka, B.~Batell, F.~Bezrukov, et~al., {\it {A
  facility to Search for Hidden Particles at the CERN SPS: the SHiP physics
  case}},  \href{http://arxiv.org/abs/1504.04855}{{\tt arXiv:1504.04855}}.

\bibitem{Anelli:2015pba}
{\bf SHiP} Collaboration, M.~Anelli et~al., {\it {A facility to Search for
  Hidden Particles (SHiP) at the CERN SPS}},
  \href{http://arxiv.org/abs/1504.04956}{{\tt arXiv:1504.04956}}.

\bibitem{Cvetic:2014nla}
G.~Cvetič, C.~Kim, and J.~Zamora-Saá, {\it {CP violation in lepton number
  violating semihadronic decays of $K, D, D_s, B, B_c$}},  {\em Phys.Rev.} {\bf
  D89} (2014) 093012, [\href{http://arxiv.org/abs/1403.2555}{{\tt
  arXiv:1403.2555}}].

\bibitem{'tHooft:1979bh}
G.~'t~Hooft, {\it {Naturalness, chiral symmetry, and spontaneous chiral
  symmetry breaking}},  {\em NATO Sci.Ser.B} {\bf 59} (1980) 135.

\end{thebibliography}\endgroup

\end{document}